# New 3,3'-(ethane-1, 2-diylidene)bis(indolin-2-one) (EBI)-based small molecule semiconductors for organic solar cells

Mylène Le Borgne,[a,b]  Jesse Quinn,[b]  Jaime Martin,[c]  Natalie Stingelin,[c] Yuning Li*[b] and Guillaume Wantz*[a]

A series of donor-acceptor-donor (D-A-D) structured small-molecule compounds, with 3,3'-(ethane-1,2-diylidene)bis(indolin-2-one) (EBI) as a novel electron acceptor building block coupled with various electron donor end-capping moieties (thiophene, bithiophene and benzofuran), were synthesized and characterized. When the fused-ring benzofuran is combined to EBI (**EBI-BF**), the molecules displayed a perfectly planar conformation and afforded the best charge tranport properties among these EBI compounds with a hole mobility of up to 0.021 cm$^2$ V$^{-1}$ s$^{-1}$. All EBI-based small molecules were used as donor material along with a PC$_{61}$BM acceptor for the fabrication of solution-processed bulk-heterojunction (BHJ) solar cells. The best performing photovoltaic devices are based on the EBI derivative using the bithiophene end-capping moiety (**EBI-2T**) with a maximum power conversion efficiency (PCE) of 1.92%, owing to the broad absorption spectra of EBI-2T and the appropriate morphology of the BHJ. With the aim of establishing a correlation between the molecular structure and the thin film morphology, differential scanning calorimetry, atomic force microscopy and X-ray diffraction analysis were performed on neat and blend films of each material.

## Introduction

Isoindigo and its derivatives have attracted much attention as π-conjugated building blocks for organic solar cells owing to their simple syntheses, strong optical absorption and appropriate electronic/energetic properties.[1,2] Since the first literature report of an isoindigo small molecule showing encouraging efficiency of 1.7%,[3] numerous isoindigo-based polymers and small molecules have been developed. While isoindigo-based polymers demonstrate a high efficiency of 7-9%,[4–9] close to 10% that is now commonly achieved with alternative π-conjugated building blocks,[10,11] only a PCE of 3-6% have so far been achieved with isoindigo-based small molecules.[12–14] Small molecules have several key advantages over polymers such as well-defined structures and a more reproducible purity; therefore, considerable efforts have been made by the community to improve isoindigo-based small molecules. Recently, Kelly *et al.* reported aza-isoindigo[15] and extended isoindigo[16] small molecule derivatives; however, OPV performances remained below 1% due to limited charge transport and unfavourable morphology. The extension of the isoindigo unit was as well realized by Welch *et al.* achieving an improved PCE of 3.7% but still limited by an undesirable morphology.[17] To establish a correlation between molecular structure and various properties such as PCE, Detrembleur *et al.* studied a series of isoindigo-based small molecules by manipulating the side chain (linear vs branched)[18] and the overall architecture (linear vs propeller-like).[19] Despite an improvement to charge transport, the PCE remained below 2%, nonetheless, the improved PCE based on the series was attributed to the more favourable charge separation which was linked to morphology. More recently, Jung *et al.* achieved a PCE of 4.33% using a diketopyrrolopyrrole flanked isoindigo small molecule. By incorporating solvent annealing an optimized PCE of 5.86% was achieved due to the formation of a bi-continuous interconnection between fullerene and the aforementioned isoindigo derivative.[20]

The low PCE of isoindigo-based small molecules is generally attributed to the limited charge transport and unfavourable morphology. The origin of the limited charge transport with isoindigo building block has already been reported to be due to the non-planar conformation influenced by steric hindrance between the 4-*H* of one indolin-2-one moiety and the nearby oxygen on the other indolin-2-one.[21] To tackle this issue, the benzene rings of isoindigo may be replaced by thiophenes, leading to a coplanar thienoisoindigo. This approach seems promising: a polymer based on thienoisoindigo showed very high hole mobilities of 14.4 cm$^2$ V$^{-1}$ s$^{-1}$. However, no improvement in PCE was observed for such thienoisoindigo derivatives, which likely is due to their high-lying HOMO level limiting the V$_{OC}$.[22–24] Another possible building block is 3,3'-(Ethane-1,2-diylidene)bis(indolin-2-one) (EBI) (Fig. 1)**,** which incorporates a butadiene bridge between the two indolin-2-one moieties and has a highly coplanar geometry due to the reduced (or eliminated) steric hindrance. EBI-based polymers absorb up to 800 nm and have a deep HOMO level of -5.38 eV providing interesting features for broad light harvesting and high open circuit voltage (V$_{OC}$).[21] Polymers comprising EBI units also showed good hole transport with mobilities of up to 0.044 cm² V⁻¹ s⁻¹ measured in a field-effect transistor configuration.

Based on these desirable optoelectronic properties of EBI-based polymers, in this study EBI was chosen as an electron withdrawing building block to develop several solution-processable donor-acceptor-donor (D-A-D- type small-molecule compounds as illustrated in Fig. 1. Specifically, three different electron rich end-capping units —i.e., thiophene (T), bithiophene (2T) and benzofuran (BF) — were attached to the EBI central core in order to manipulate the optoelectronic properties, solubility, solution-processability, and solid state structure formation of the resulting materials. The charge transport and photovoltaic performances of these new compounds were evaluated in OTFTs and BHJ solar cells. To address the morphology issue commonly displayed by isoindigo-based small molecules, an extensive study of the bulk heterojunction formation was conducted using phase diagram, atomic force microscopy, and X-ray diffraction.

## Results and discussion

### Synthesis and characterization

The synthetic routes of the new D-A-D type EBI small-molecules, EBI-T, EBI-BF, and EBI-TT, are described in Scheme 1. The end capping donor units were expected to be able to extend the π-conjugation, reduce the band gap, and tune the frontier energy levels of these molecules, which are important for improving the charge transport and the matching of these materials with the acceptor (PCBM) for OPVs. Thiophene (T) benzofuran (BF), and bithiophene (TT) are three representative electron-donating building blocks for high performance small molecule and polymer semiconductors for OPVs. These building blocks were expected to result in different π-conjugation length and different energy levels, e.g., T vs. TT and TT vs. BF, which would provide useful insight in to the structure-property relationship of these EBI molecules. Additionally, for the EBI building block, the non-substituted five membered heteroaryl donor units that are adjacent to the benzene ring of EBI would be ideal for achieving a high coplanarity for higher mobility and thus a smaller band gap for improved sunlight harvesting. (3Z)-6-Bromo-3-((Z)-2-(6-bromo-2-oxoindolin-3-ylidene)ethylidene)indolin-2-one (1), the key starting material to form all EBI derivatives, was prepared using published procedures.[21,25] The synthesis consists of mixing 6-bromoindole-2,3-dione with propionic anhydride and pyridine under reflux. Subsequently, 2-butyloctyl chains were introduced onto the nitrogen atoms of the indoline-2-one moieties to obtain the intermediate compound, 2a. EBI compounds capped with thiophene (EBI-T) and benzofuran (EBI-BF-C₁₂) were prepared through the Stille coupling reactions of 2a with tributyl(thiophen-2-yl)stannane and trimethyl(benzofuran-2-yl)stannane, respectively, at 90 and 110 °C for 60 h. The catalytic system used for the Stille coupling reactions was a mixture of Pd₂(dba)₃ and P(o-tolyl)₃. Tributyl(thiophen-2-yl)stannane was obtained commercially whereas trimethyl(benzofuran-2-yl)stannane was produced by reacting benzofuran with n-butyllithium and then trimethyl tin chloride. EBI-T and EBI-BF-C₁₂ were obtained with moderate yields of 74% and 55%, respectively, after purification by column chromatography. The solubility of EBI-BF-C₁₂ was found to be notably lower than that of EBI-T in chloroform, indicating stronger intermolecular interactions of the former. To improve the solubility of EBI-BF, longer 2-hexyldecyl chains were attached and the resulting EBI-BF-C₁₆ was prepared in 65% yield following a similar procedure used for preparing EBI-BF-C₁₂. As for EBI-2T, the EBI compound capped with bithiophene, a Suzuki coupling reaction was carried out between 2a and a commercial 4,4,5,5-tetramethyl-2-(5-(thiophen-2-yl)thiophen-2-yl)-1,3,2-dioxaborolane in presence of a catalyst, Pd(PPh₃)₄, and a base, K₂CO₃ (2M), at 85 °C for 24 h. EBI-2T was obtained with a yield of 67% after purification by column chromatography. All four EBI compounds were fully characterized using ¹H NMR, ¹³C NMR, and field desorption mass spectrometry.

Fig. 1 Chemical structures of the EBI-based molecules investigated.

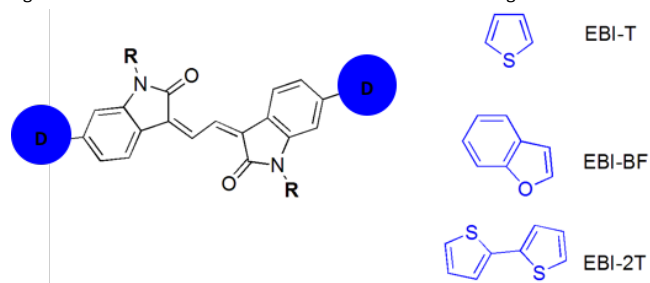

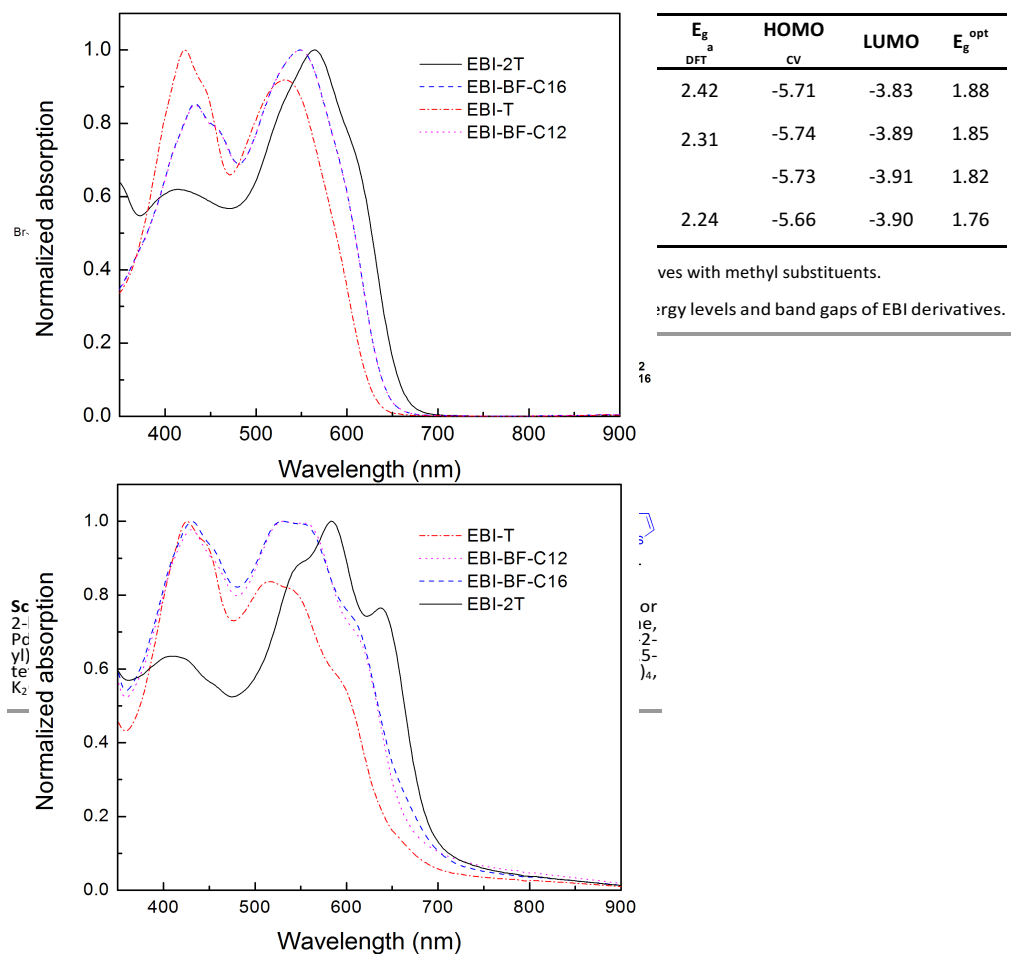

| $E_g^{DFT}$ | HOMO$_{CV}$ | LUMO | $E_g^{opt}$ |
|---|---|---|---|
| 2.42 | -5.71 | -3.83 | 1.88 |
| 2.31 | -5.74 | -3.89 | 1.85 |
| | -5.73 | -3.91 | 1.82 |
| 2.24 | -5.66 | -3.90 | 1.76 |

ves with methyl substituents.

rgy levels and band gaps of EBI derivatives.

**Fig. 2** UV-Visible absorption spectra of EBI derivatives in solution (top) and in film (bottom).

## Optical properties

The UV-Vis absorption spectra of the EBI compounds were recorded in chloroform solutions and in thin films. They are shown in Fig. 2. In solution, all EBI compounds exhibited two-band profiles. The 300-500 nm band corresponds to the π-π* transition within the electron rich end-capping units whereas the 500 − 700 nm band can be assigned to a band resulting from intramolecular charge transfer (ICT) from the electron-rich end units (D) to the electron deficient EBI unit (A). EBI-T showed a weaker absorption peak in the second band than in the first band, suggesting that ICT between thiophene and EBI is less efficient. In comparison, EBI-2T and EBI-BF compounds presented stronger ICT. The wavelengths of maximum absorption ($\lambda_{max}$) for EBI-2T and EBI-BF is found at 549 nm and 565 nm, i.e. red-shifted compared to that of EBI-T ($\lambda_{max}$ = 531 nm). From solution to solid state, only $\lambda_{max}$ of EBI-2T red-shifted; yet an additional shoulder appeared for all EBI compounds, which we attribute to intermolecular interactions that establish in the solid state. It is furthermore noteworthy that the two EBI-BF compounds have similar UV-Vis spectra in both solution and solid-state suggesting that the length of the alkyl chains do not impact their aggregation in the solid state. The optical band gaps of EBI-T, EBI-BF and EBI-2T, estimated from the absorption edge of the film, is 1.88 eV, 1.85 eV, and 1.76 eV, respectively. EBI-2T is the most propitious for harvesting sunlight.

## Energy levels

To depict the electronic structures of the EBI compounds, computer simulations based on density functional theory were conducted using Gaussian B3LYP and 6-31G basis set on three model compounds with methyl substituents for the convenience of computation. The optimized backbone conformations are shown in Fig. 3. They confirm the high planarity of the EBI core units. As for the conformations of the entire molecule, EBI-T and EBI-2T exhibit dihedral angles of, respectively, 26 ° and 22 ° between the EBI and thiophene units whereas EBI-BF has a fully planar conformation. To gain further information concerning the impact of electron-rich end-capping units on the energy levels of a given compound, the LUMO and HOMO electron density distributions of these model compounds were calculated. These are shown in Fig. 3b and 3c, respectively. The HOMO electron densities of EBI-T, EBI-BF and EBI-2T are distributed along the entire backbones whereas the LUMO electron densities are localized on the EBI units and thus the donor units should mainly impact the HOMO levels of these molecules. The calculated HOMO and LUMO levels are listed in Table 1. EBI-T has the largest band gap due to its shortest π-conjugation length consistent with the UV-Vis data. EBI-BF has the deepest HOMO indicating the weakest electron-donating effect of benzofuran among the three end-capping units.

The electrochemical properties were investigated by cyclic voltammetry in 0.1 M tetrabutylammonium hexafluorophosphate **in** acetonitrile using two platinum electrodes as counter and working electrodes and an Ag/AgCl (0.1 M) electrode as the reference electrode. The ferrocene/ferrocenium couple, which has a HOMO level of -4.8 eV, was used as an energy level reference.[26] Films of EBI compounds were deposited by drop-casting on the working electrode. HOMO levels were calculated from their onset oxidation potentials to be -5.71 eV for EBI-T, -5.73 eV for EBI-BF and -5.66 eV for EBI-2T. These high HOMO energy levels are desirable for achieving a high $V_{OC}$ of > 1 V according to the Scharber equation.[27] Since the reduction peaks were very weak, the LUMO levels were obtained by adding the optical band gaps to the HOMO levels to be -3.83 eV for EBI-T, -3.89 eV for EBI-BF and -3.90 eV for EBI-2T. The offsets between the LUMOs of these EBI compounds and $PC_{61}BM$ (-4.3 eV)[27] are larger than the minimum value of ~0.3 eV required for the electron to efficiently transfer from the donor molecules to $PC_{61}BM$. By using the energy levels of EBI derivatives on the Scharber diagram,[27] PCE of 7%, 8% and 9% can be expected for EBI-T, EBI-BF and EBI-2T, respectively, providing further evidence of the promise of these compounds for potentially leading to high performance OSCs.

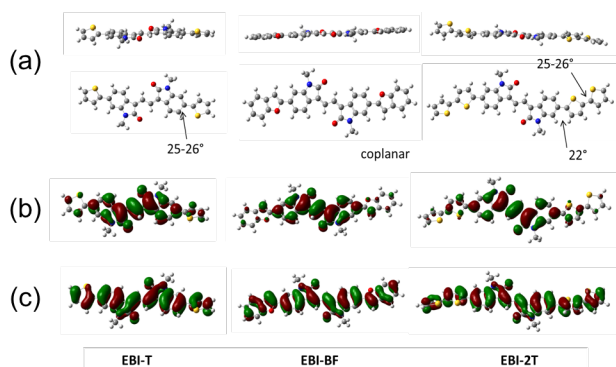

**Fig. 3** DFT calculations: lowest energy conformation (a) and electron density representation for LUMO (b) and HOMO (c) level, respectively.

**Charge transport properties**

The charge transport properties of our EBI derivatives were evaluated in bottom-gate bottom-contact organic thin film transistors (OTFTs). The devices were fabricated using heavily n-doped Si/$SiO_2$ wafer substrates with pre-patterned gold source-drain electrode pairs. A chloroform solution of 10 mg $mL^{-1}$ of the EBI compound was spin-coated onto the substrate to finalise the devices. OTFTs based on these EBI compounds showed typical hole-transport characteristics. The spin-coating speeds and thermal annealing conditions were varied to control the molecular packing of each EBI compound. An as-cast thin film of EBI-T, deposited at 2500 rpm, displayed an average saturation hole mobility ($\mu_h$) of $6.4 \times 10^{-5}$ $cm^2$ $V^{-1}$ $s^{-1}$ and a maximum mobility of $2.3 \times 10^{-4}$ $cm^2$ $V^{-1}$ $s^{-1}$. For EBI-2T, a slow deposition speed which provided slow drying, led to larger crystals (see ESI†) resulting in a saturation hole mobility to up to $1.5 \times 10^{-3}$ $cm^2$ $V^{-1}$ $s^{-1}$ (Table 2). Relatively high on/off current ratios of ~$10^5$ were achieved along with a small threshold voltage of 4 V.

Without any post treatment, EBI-BF derivatives exhibited limited saturation hole mobilities of $4.5 \times 10^{-4}$ $cm^2$ $V^{-1}$ $s^{-1}$ for the EBI-BF-$C_{12}$ with shorter alkyl chains and $2.9 \times 10^{-5}$ $cm^2$ $V^{-1}$ $s^{-1}$ for the EBI-BF-$C_{16}$ with longer alkyl chains. Fig. 4 summarizes the hole mobilities measured for OTFTs based on films of these two EBI-BF compounds, as-cast and after successive annealing at 100, 150, and 200 °C for 10 minutes. The saturation hole mobilities of EBI-BF-$C_{12}$ and EBI-BF-$C_{16}$ increased to $4.4 \times 10^{-3}$ $cm^2$ $V^{-1}$ $s^{-1}$ and 0.02 $cm^2$ $V^{-1}$ $s^{-1}$ upon thermal annealing at 100 °C and 150 °C, respectively. Under the optimal thermal annealing condition, EBI-BF-$C_{16}$ displayed better charge transport than EBI-BF-$C_{12}$. In addition, the leakage current was reduced and the on/off ratio was improved to $10^6$ compared to those of EBI-BF-$C_{12}$.

The effect of thermal annealing was further investigated by comparing the UV-Vis absorption spectra of the as-cast and annealed EBI-BF films (see ESI†). Upon annealing at 100 °C, a more pronounced vibronic structure is observed for EBI-BF-$C_{16}$ along with an increase of the peak intensity at 650 nm. For EBI-BF-$C_{12}$, an additional shoulder appears at 653 nm after an annealing at 140 °C which indicates a change in the electronic coupling, possibly resulting from a different molecular packing that favours charge transport. The higher mobility observed for the EBI-BF compounds compared to those of the other two EBI compounds may be due to their coplanar conformations allowing closer π-π packing. Most importantly, the hole mobilities of both compounds seem sufficiently high that they should sustain necessary charge extraction in OSCs made of these materials to reach high fill factors.[28] In contrast, the charge transport properties of EBI-T (in the range of $10^{-4}$ $cm^2\ V^{-1}\ s^{-1}$), may be too low for OSC applications and could lead to large carrier sweep-out times and therefore bimolecular recombination; while EBI-2T may just sustain a sufficiently high hole mobility as long as the addition of the acceptor does not affect its molecular packing drastically.

| Speed deposition | $\mu_{h\ average}$ * (L=5μm) ($cm^2\ V^{-1}\ s^{-1}$) | $\mu_{h\ max}$ (L=5μm) ($cm^2\ V^{-1}\ s^{-1}$) | $\mu_{h\ average}$ * (L=10μm) ($cm^2\ V^{-1}\ s^{-1}$) | $\mu_{h\ max}$ (L=10μm) ($cm^2\ V^{-1}\ s^{-1}$) |
|---|---|---|---|---|
| **2 500 rpm** | $1.52 \times 10^{-3}$ +/- $1.15 \times 10^{-3}$ | $3.17 \times 10^{-3}$ | $1.67 \times 10^{-3}$ +/- $1.09 \times 10^{-3}$ | $2.56 \times 10^{-3}$ |
| **5 000 rpm** | $4.43 \times 10^{-5}$ +/- $5.5 \times 10^{-6}$ | $5 \times 10^{-5}$ | $5.3 \times 10^{-5}$ +/- $5.13 \times 10^{-5}$ | $1.26 \times 10^{-4}$ |
| **10 000 rpm** | $2.52 \times 10^{-5}$ +/- $5.5 \times 10^{-6}$ | $3.1 \times 10^{-5}$ | $1.92 \times 10^{-5}$ +/- $1.9 \times 10^{-6}$ | $2.18 \times 10^{-5}$ |

**Table 2** Hole mobilities measured in EBI-2T based OTFTs as a function of spin-casting conditions W = 1 cm; L = 5 or 10 μm; $V_{DS}$ = -40 V.

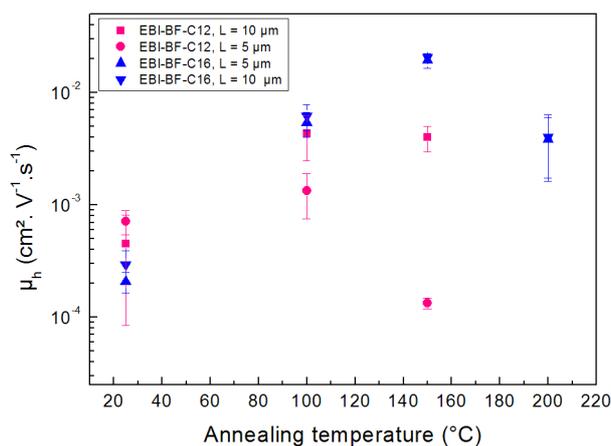

**Fig. 4** Hole mobilities of EBI-BF-$C_{16}$ and EBI-BF-$C_{12}$ in OTFTs after various thermal annealing at 100, 150, and 200 °C for 10 min. W = 1 cm; L = 5 or 10 μm; $V_s$ = -40 V.

**Photovoltaic properties**

The photovoltaic performance of EBI-T, EBI-BF-$C_{12}$ and EBI-2T was first investigated in bi-layer OSCs with the following architecture: ITO/PEDOT:PSS/EBI/$C_{61}$/Ca/Al. EBI compounds were deposited by spin-coating chloroform solutions to form films with thicknesses of ~30 nm. Among the prepared devices, EBI-2T exhibited the highest $J_{SC}$ possibly due to its broad absorption spectrum. EBI-T did not show any photovoltaic effect. We attribute this to the low charge transport properties of this material that can lead to pronounced recombination, as alluded to already above.

To gain further information on the bulk heterojunction (BHJ) formation of our EBI-based compounds, we first elucidated their miscibility and general phase behaviour with the acceptor PC$_{61}$BM. For this purpose, we conducted DSC measurements on blend films of various EBI/PC$_{61}$BM ratios, drop-cast from chloroform solutions. Phase diagrams were established by plotting the end values of the endotherms observed in the first heating thermograms against the weight percentage (wt%) of the donor in the blend samples (Fig. 5a-b), as described previously.[29] Both blend systems based on these two EBI-compounds display a typical eutectic phase behaviour. The eutectic compositions and the eutectic temperatures are, respectively, 55% and ~210 °C for EBT-T and 50% and ~245 °C for EBI-2T. This phase behaviour suggests that both EBI-T and EBI-2T are miscible to a certain extent with PC$_{61}$BM. It was more challenging to analyse the two EBI-BF compounds because both EBI-BF-$C_{12}$ and EBI-BF-$C_{16}$ feature a rather low solubility in organic solvents. Hence, merely films of poor quality could be prepared with EBI-BF-$C_{12}$ and blend films with PC$_{61}$BM. We therefore only assessed the phase behaviour of EBI-BF-$C_{16}$ and EBI-BF-$C_{16}$:PC$_{61}$BM binaries. Fig. 5c shows the first DSC heating thermograms of these blend systems. A small melting point depression can be observed for both the donor and the acceptor; however, no clear eutectic temperature can be identified. Based on our data, we

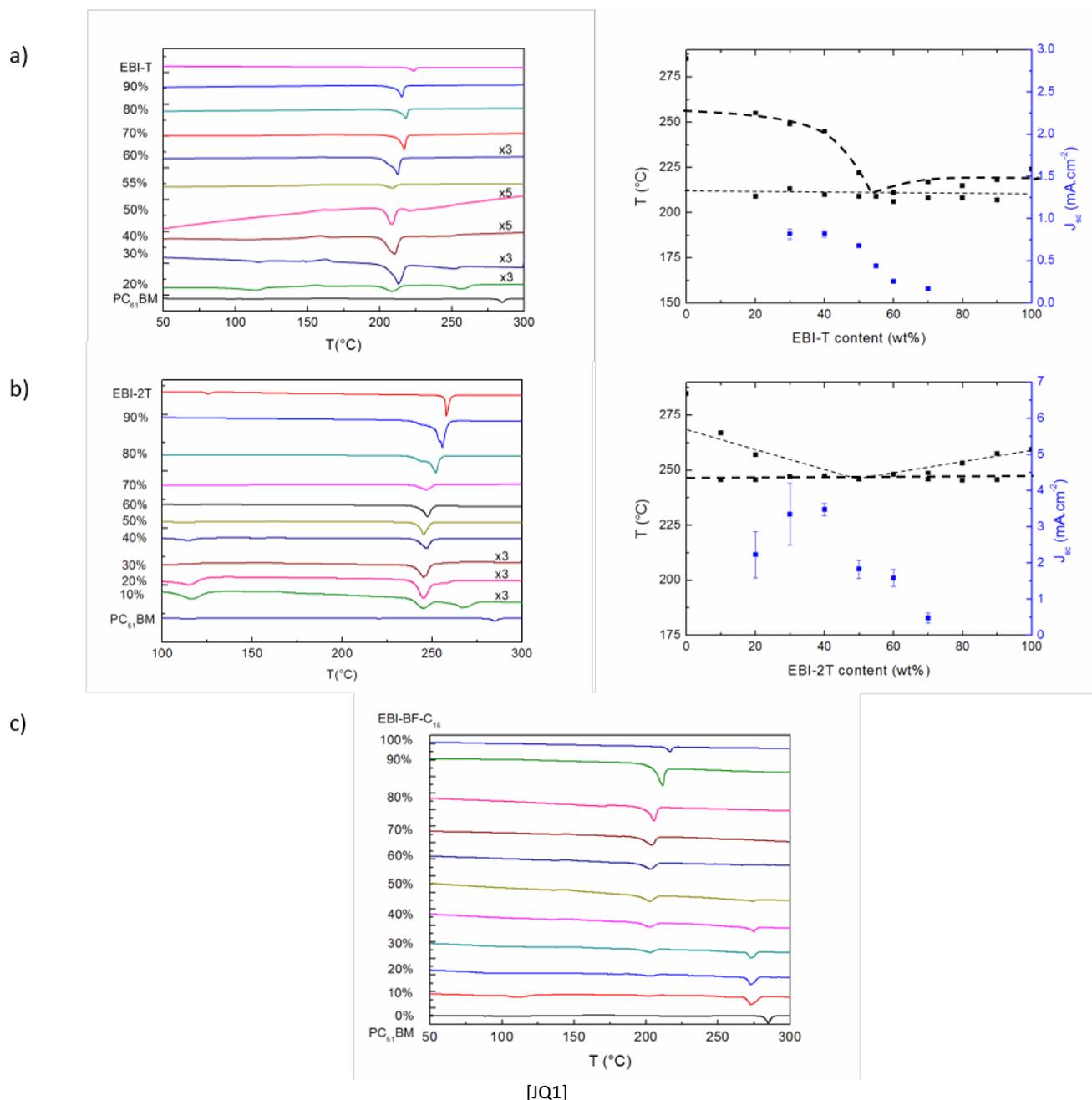

[JQ1]

**Fig. 5** DSC first-heating thermograms and phase diagrams of EBI-T:PC$_{61}$BM blends (a) and EBI-2T:PC$_{61}$BM blends (b). For comparison, corresponding data for EBI-BF-C$_{16}$:PC$_{61}$BM blends are also shown (c).

cannot exclude that the two compounds feature a liquid-liquid phase separation-regime at certain compositions. Clear is, that the mutual miscibility of this binary is lower than those comprising EBT-T and EBI-2T.

Since a direct correlation has previously been made between the optimum device composition and the eutectic phase behaviour,[29–33] we prepared devices with blends of a donor:acceptor ratio around the eutectic composition: i.e. we produced devices with binaries comprising 30% to 70 wt% of EBI-T (or EBI-2T EBI) using a standard BHJ OSC architecture of ITO/PEDOT:PSS/EBI:PC$_{61}$BM/Ca/Al. The EBI/PC$_{61}$BM ratio mainly affected J$_{SC}$ (Fig. 5b and d). Similar to the results reported in the literature,[29–33] the maximum J$_{SC}$ for both EBI-T and EBI-2T OSCs was found for slightly hypoeutectic blends (40 wt% donor; i.e. a slightly fullerene-rich blend). This is a composition for which one can expect a good trade-off between charge separation and charge transport[29–33] which led for the EBI-2T systems to an average PCE of 1.65%, with the highest value reaching 1.92% (

| | Ratio | JSC | VOC | FF | PCE (average/max) |
|---|---|---|---|---|---|
| **EBI-T** | 2:3 | 0.82 | 0.73 | 0.30 | 0.18% /0.21% |
| **EBI-2T** | 2:3 | 5.51 | 0.87 | 0.34 | 1.65%/ 1.92% |

|  | Ratio | $J_{SC}$ | $V_{OC}$ |
|---|---|---|---|
| **EBI-T** | 2:3 | 0.82 | 0.73 |
| **EBI-2T** | 2:3 | 5.51 | 0.87 |
| **EBI-BF-C$_{16}$\*[JQ2]** | 1:1 | 0.87 | 0.23 |

**Table 3** BHJ performances of EBI-T, EBI-2T a~~

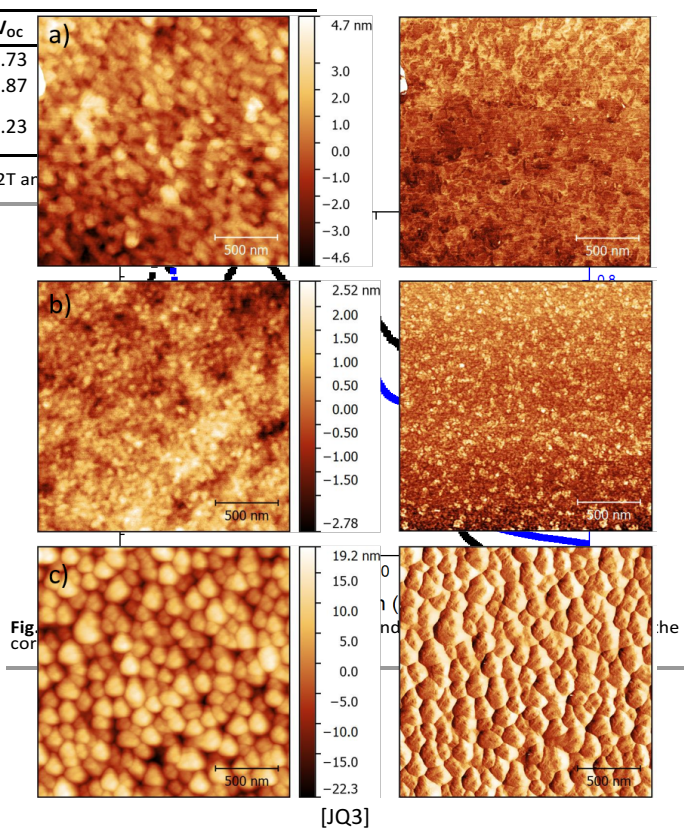

**Fig. 7** AFM height (left) and phase (right) of EBI-T:PC$_{61}$BM blend in a 2:3 ratio (a), EBI-2T:PC$_{61}$BM blend in a 2:3 ratio (b) and EBI-BF-C$_{16}$:PC$_{61}$BM blend in a 1:1 ratio (c).

[JQ3]

| **EBI-BF-C16\*** | 1:1 | 0.87 | 0.27 | 0.06%/ 0.08% |
|---|---|---|---|---|

). As for the bi-layer devices, EBI-BF-based blends showed a poor photovoltaic performance also in BHJ architectures: we find an average PCE of only 0.06%, even when the donor:acceptor ratio was changed or the films were annealed. The lower PCE of these devices compared to the other EBI-blend systems was mainly due to the drastically decreased $J_{SC}$ and $V_{OC}$, which we attribute to the low miscibility of the two components, drastically limiting charge generation.[34]

As EBI-2T exhibited the highest $J_{SC}$, we measured the quantum efficiency (EQE) for our best cell prepared with a EBI-2T:PC$_{61}$BM. As can be seen from **¡Error! No se encuentra el origen de la referencia.**, an EQE of 30 − 35% can be recoded for an incident radiation with a wavelength between 350 − 650 nm, indicating that photons absorbed by EBI-2T participated in the photovoltaic process. By integrating the spectrum, a $J_{SC}$ value of 5.4 mA cm$^{-2}$ was calculated which is consistent with AM1.5 IV values.

**Solid-state microstructure**

For BHJ OSCs, the device performance is highly dependent on the solid-state microstructure and phase morphology of the active layer which is ideally composed of three phases: an intermixed donor/acceptor phase, a pure crystalline donor phase and a pure crystalline acceptor phases. The intermixed network promotes efficient and fast charge separation[34–36] whereas pure crystalline domains ensure charge transport to the electrodes.[37] Hence, indication of the mutual miscibility of the donor and the acceptor, as obtained via establishing phase diagrams is important. Further insight can be gained from atomic force microscopy (AFM) and X-ray diffraction (XRD) on the active layer of the optimal devices of each material.

The height and phase images of each EBI active layer are shown in Fig. 7. The height images of EBI-T and EBI-2T exhibited relatively smooth films with a square root roughness of less than 4 nm, which is beneficial for forming good interfaces with the calcium cathode. The phase images reveal a relatively small domain structure for both systems, which is in accord with the phase behaviour we deduce from differential scanning calorimetry. In stark contrast, both height and phase images of the EBI-BF-C$_{16}$ samples show large grains of 100 − 250 nm in size. There seems only one material to be present on the top surface of these blend films suggesting a vertical phase segregation of the materials. This may be in accord with the relatively low roughness of only 6.3 nm that we observe for these films where we would expect a more pronounced phase separation between the donor and acceptor based on our thermal analysis data. Such a vertical phase separation , would also hinder charge extraction from a BHJ OSC, especially in case this top-layer is EBI-BF-$C_{16}$ rich, which would contribute to the poor photovoltaic performance of EBI-BF-$C_{16}$:$PC_{61}$BM blends.[38,39]

The question remains why EBI-2T devices outperform the EBI-T OSCs. For this, we performed XRD measurements on neat thin films spin-coated from chloroform solutions without thermal annealing. As shown in Fig. 8, EBI-BF-$C_{12}$, EBI-BF-$C_{16}$ and EBI-2T display well defined and intense reflexion peaks indicative of good molecular order. Based on the peak position, the *d*-spacing can be calculated using the Bragg's equation to be 2.0 nm for EBI-BF-$C_{12}$, 2.3 nm for EBI-BF-$C_{16}$ and 1.6 nm for EBI-2T. In contrast, EBI-T showed only a weak broad peak, indicative of a relatively poor crystallinity, indicating that devices based on this donor may be limited also by charge extraction. In contrast to the EBI-BF-$C_{16}$:$PC_{61}$BM devices, charge extraction seems low because of the low molecular order of the donor-rich phase limiting the hole charge transport (in agreement with our TFT measurements).

The crystallite domain sizes were calculated using Scherrer equation: D= 0.9λ/βcosθ, where λ is the X-ray wavelength, β is the full width at half maximum in radians, and θ is the scattering angle. The crystallite domain sizes of neat EBI-BF-$C_{12}$, EBI-BF-$C_{16}$ and EBI-2T are 32.2 nm, 24.0 nm and 39.4 nm, respectively. The crystal sizes of the blend with a donor: acceptor ratio of 1:1 for EBI-BF derivatives and 2:3 for EBI-2T were also calculated. We find a decreased domain size of 33.8 nm for the EBI-2T blend compared to the neat EBI-2T film. This is in agreement with the observation made in thermal analysis that EBI-2T and $PC_{61}$BM are miscible to form intermixed domains hindering the crystallization of the donor. For EBI-BF-$C_{16}$:fullerene blends, we observed a weakened peak at 2θ ∼ 4 º that can be attributed to the molecule. Additionally, a new peak of unknown origin appeared at 2θ ∼ 6º. Potentially, this new peak may be due to the presence of $PC_{61}$BM crystals or the EBI-BF-$C_{16}$ crystals having a different orientation or formation crystals of EBI-BF-$C_{16}$ with a different structure

Summarising our findings from AFM and XRD analyses: the EBI-2T:$PC_{61}$BM system seems to display a three-phase morphology. This is consistent with its phase behaviour deduced from DSC and the relatively good OSC performance of this donor:acceptor blend systems. For the EBI-T:$PC_{61}$BM system, small domains are observed in AFM; however the neat donor as well as the corresponding blend with $PC_{61}$BM seem to be of a lower large-scale molecular order as deduced from XRD – although relatively clear endotherms are observed in thermal analysis.

The poor crystallinity of EBI-T seems to limit charge transport and, thus, charge extraction, which could explain its low performance in OSC devices. In stark contrast to the EBI-T and EBI-2T blends, the EBI-BF-$C_{16}$:$PC_{61}$BM blend films exhibit large grains on the top surface of blend films which may indicate a preferred vertical phase separation driven by the low compatibility of the two compounds, possibly induced by the planar conformation of EBI-BF-$C_{16}$ leading to an unfavourable entropy of mixing.[40] Our series of EBI derivatives provides some important materials design criteria.

## Conclusions

3,3'-(Ethane-1,2-diylidene)bis(indolin-2-one) was employed as an electron accepting core to design four new D-A-D type electron donor semiconductors, EBI-T, EBI-2T, and EBT-BF (-$C_{12}$ and −$C_{16}$), by incorporating various electron rich end-capping units: thiophene, bithiophene and benzofuran. OSCs based on EBI-2T:$PC_{61}$BM system showed the highest performances, with a maximum PCE of 1.92%. The limited performance of EBI-T based OSCs seems to be caused by the low crystallinity and thus poor charge transport of EBI-T, while a low miscibility that leads to a coarse phase separation (and likely to a vertical phase separation), limiting both charge separation and charge extraction, undesirably affects the OSC performance of EBI-BF-$C_{16}$-based blends. Most importantly, we show that the end-capping unit has a significant effect on the energy levels, band gaps, charge transport property, morphology, and the ultimate photovoltaic performance of the EBI compounds, giving us clear materials design guidelines. Finally, we also like to highlight that the materials presented here are the first EBI-based small molecule semiconductors for solution-processable bulk heterojunction solar cells reported so far, opening a new range of materials to be developed. Further exploration of chemical derivatives is however necessary to improve their solar performances.

## Experimental section

### Materials and methods

All reagents and chemicals were purchased from commercial sources and used without further purification. (3*Z*)-6-Bromo-3-((*Z*)-2-(6-bromo-2-oxoindolin-3-ylidene)ethylidene)indolin-2-one (**1**) was prepared as described in the literature[21]. All [1]H NMR and [13]C NMR spectra were recorded on a 300 MHz Bruker NMR using

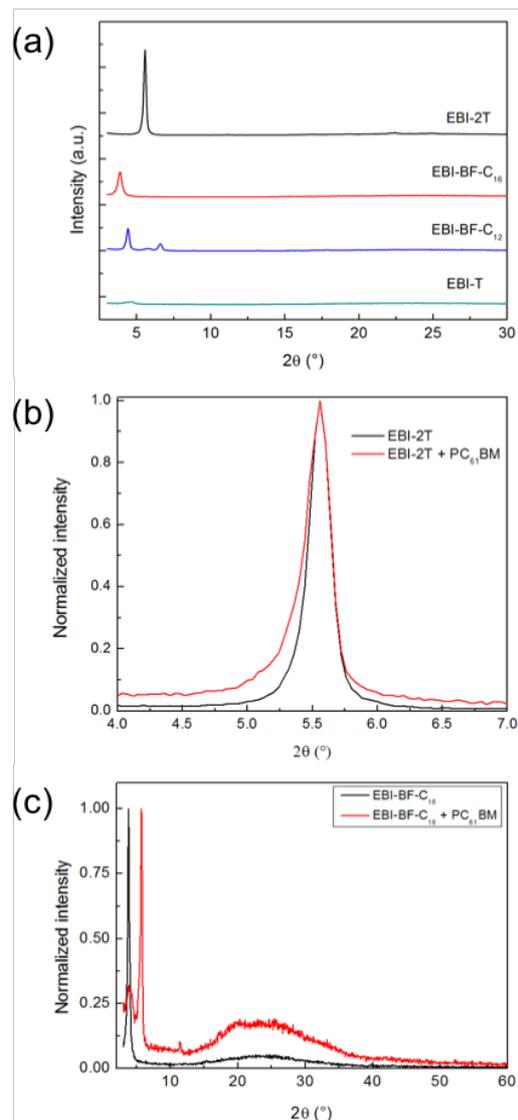

**Fig. 8** XRD spectrum of EBI derivatives in neat film (a) and in blend film for EBI-2T:PC$_{61}$BM 2:3 (b) and EBI-BF-C$_{16}$:PC$_{61}$BM 1:1 (c).

CDCl$_3$ or TCE-d$_2$ as solvents and their peak as a reference. High resolution field desorption technique mass spectrometry was used for all final EBI compounds and high resolution electrospray technique mass spectrometry for the intermediate compounds. Cyclic voltammetry (CV) measurements were conducted using a DY2000EN electrochemical workstation in a 0.1 M tetrabutylammonium hexafluorophosphate acetonitrile electrolyte solution at room temperature at a scan rate of 50 mV s$^{-1}$. The working electrode and counter electrode were platinum electrodes and the reference electrode was Ag/AgCl (0.1 M) electrode. The ferrocene/ferrocenium (Fc/Fc$^+$) couple was used as the reference, which has a JOMO energy of -4.8 eV. [*Adv. Mater.* 1995, 7, 551] The thin films on the Pt disk, formed by drop-casting solutions of the EBI compounds in chloroform, were used as the working electrode. The following equation was used to calculate the HOMO energy level: E$_{HOMO}$ = - (E$^{ox}_{onset}$ − (E$_{Fc}$ -4.8)) eV. AFM imaging was carried out at room temperature using an AFM Nanoman from Bruker Instrument with Nanoscope 5 controller. Images were obtained in a tapping mode using silicon tips (PointProbe® Plus AFM-probe, Nanosensors, Switzerland). Differential scanning calorimetry (DSC) was performed on a Mettler−Toledo DSC 1 Stare system at a heating/cooling rate of 10 °C min$^{-1}$. The molecular packing was characterized by grazing incidence with θ constant (<1.5°). X-ray diffraction (XRD, D8 Advance diffractometer) using the Cu Kα radiation. θ/2θ scans were performed to the spin-coated films at room temperature.

**Synthesis**

**(3*Z*)-6-Bromo-3-((*Z*)-2-(6-bromo-2-oxoindolin-3-ylidene) ethylidene)-indolin-2-one (1):**

A reaction of a mixture of 6-bromoindole-2,3-dione (15 g, 0.066 mol, 1 eq), propionic anhydride (50 ml) and pyridine (17 ml) was heated to reflux in a 100 ml three-necked round-bottom flask under argon for 30 min. After cooling, the dark red needles were separated, washed with acetone and ether several times and dried under vacuum at 100°C. After washing three times with chloroform, 4.6 g of (3Z)-6-Bromo-3-((Z)-2-(6-bromo-2-oxo-1-propionylindolin-3-ylidene)ethylidene)-1-propionylindolin-2-one intermediate were obtained with a yield of 25%. Then, 4 g of the intermediate was further reacted with 0.8 g of KOH in 100 ml of ethanol under reflux for 20 minutes. The dark blue solid was separated, washed with acetone and ether. The potassium derivatives were neutralized with diluted HCl (0.20 M) in ethanol and washed with water. 2.73 g of final product was obtained affording 85% yield.

$^1$H NMR (DMSO, 300MHz) δ (ppm) 10.81 (d, J=21.7 Hz, 1H), 8.65 (d, J=13.1 Hz, 1H), 8.07 (dd, J=15.0, 10.8 Hz, 2H),7.97 (d, J=8.2 Hz, 1H), 7.18 (t, J=7.6 Hz, 2H), 7.01 (d, J=12.9 Hz, 2H)

### (3Z)-6-Bromo-3-((Z)-2-(6-bromo-1-(2-butyloctyl)-2-oxoindolin-3-ylidene)ethylidene)-1-(2-butyloctyl)indolin-2-one (2a):

In a two-necked round bottom flask, **1** (1.5 g, 3.4 mmol, 1 eq), potassium carbonate (1.81 g, 13.1 mmol, 3.9 eq) and 14 ml of DMF were introduced. 2-Butyloctyl bromide (2.5 g, 10.0 mmol, 2.9 eq) was then added. The reaction was stirred at 70°C under argon for 20 h. 35 ml of deionized water were added and the product was extracted with dichloromethane (DCM). The organic phase was dried with sulphate sodium and DCM was evaporated under vacuum. The product was further purified on silica gel chromatography column using a mixture of Hexane and DCM (3:2) as eluent. The red solid was crystallized in isopropanol three times to get red needles. 612.8 mg of **2a** was obtained with a yield of 23.0%.

$^1$H NMR (CDCl3, 300MHz): δ 8.94 (s, 2H), 7.47 (d, J = 8.2 Hz, 2H), 7.16 (d, J = 8.1 Hz, 2H), 6.91 (s, 2H), 3.58 (d, J = 7.2 Hz, 4H), 1.86 (s, 2H), 1.43 – 1.13 (m, 37H), 1.11 – 0.61 (m, 13H). $^{13}$C NMR (CDCl₃, 300MHz) δ 167.18, 144.52, 130.28, 129.00, 124.98, 124.40, 122.36, 122.09, 112.06, 44.41, 36.25, 31.78, 31.58, 31.30, 29.62, 28.65, 26.40, 23.03, 22.62, 14.07, 14.03. HR-MS (ESI): calculated for C₄₂H₅₈O₂N₂Br₂: 781.29514; found (M+H)⁺: m/z 781.29378.

### (3Z)-6-Bromo-3-((Z)-2-(6-bromo-1-(2-hexyldecyl)-2-oxoindolin-3-ylidene)ethylidene)-1-(2-hexyldecyl)indolin-2-one (2b):

In a two-necked round bottom flask, **1** (1.35 g, 0.003 mol, 1 eq), potassium carbonate (1.63 g, 0.0118 mol, 3.9 eq) and 13 ml of DMF were introduced. 2-hexyldecyl iodide (3.17 g, 0.009 mol, 3 eq) was then added. The reaction was stirred at 70°C under argon for 20 hours. 40 ml of deionized water were added and the product was extracted with DCM. The organic phase was dried with sulfate sodium and DCM was evaporated under vacuum. The product was further purified on silica gel chromatography column using a mixture of Hexane and DCM (65:35) as eluent. The red solid was crystallized in isopropanol three times to get red needles. 530 mg of **2b** was obtained giving a yield of 20.0%.

$^1$H NMR (CDCl₃, 300MHz) δ (ppm) 8.94 (s, 2H), 7.47 (d, J= 7.47 Hz, 2H), 7.16 (d, J=7.16 Hz, 2H), 6.91 (s, 2H), 3.57 (d, J=7.3 Hz, 4H), 1.86 (s, 2H), 1.47-1.11 (m, 49H), 0.95-0.78 (m, 12H). $^{13}$C NMR (CDCl₃, 300MHz) δ (ppm) 167.19, 144.48, 130.27, 129.01, 124.98, 124.39, 122.35, 122.06, 112.07, 44.39, 36.20, 31.87, 31.78, 31.57, 31.55, 29.95, 29.63, 29.52, 29.28, 26.42, 26.39, 22.66, 22.63, 14.11, 14.09. HR-MS: calculated for C₅₀H₇₄O₂N₂Br₂: 893.41948; found (M+H)⁺: m/z 893.41898

### (3Z)-1-(2-butyloctyl)-3-((Z)-2-(1-(2-butyloctyl)-2-oxo-6-(thiophen-2-yl)indolin-3-ylidene)ethylidene)-6-(thiophen-2-yl)indolin-2-one (EBI-T):

**2a** (0.70 g, 0.9 mmol, 1 eq) and tributyl(thiophen-2-yl)stannane (0.57 ml, 1.8 mmol, 2 eq) were placed in a 50 ml dried flask and then, the flask was evacuated and back-filled with argon three times. Next, 30 ml of anhydrous toluene was poured into the flask. A mixture of Pd₂(dba)₃ (0.019 g, 0.018 mmol, 0.02 eq) and P(o-tolyl)₃ (38.5 mg) which were separately introduced in 1 ml of anhydrous toluene were injected into the flask. The mixture was heated at 90°C under argon and stirred for 60 hours. Toluene was evaporated. The compound was purified on silica gel chromatography column with hexane and DCM (60:40) as eluent. The product was crystallized in isopropanol three times. 0.53 g of EBI-T was obtained giving a yield of 74.6%.

$^1$H NMR (CDCl₃, 300MHz) δ 8.96 (s, 2H), 7.61 (d, J=7.9 Hz, 2H), 7.35 (dd, J1= 3.6 Hz, J2= 1.0 Hz, 2H), 7.32 (dd, J1= 5.1 Hz, J2= 1.0 Hz), 7.27 (dd, J1=7.9 Hz, J2= 1.5 Hz, 2H), 7.09 (dd, J1= 5.1 Hz, J2=3.6 Hz, 2H), 6.97 (d, J=1.1 Hz, 2H), 3.65 (d, J =7.4 Hz, 4H), 1.91 (s, 2H), 1.37-1.12 (m, 32H), 1.02-0.69 (m, 12H). $^{13}$C NMR (CDCl₃, 300MHz) δ 167.66, 144.27, 143.98, 136.31, 130.27, 128.59, 128.21, 125.55, 123.69, 122.61, 121.70, 119.57, 105.80, 44.17, 36.42, 31.82, 31.76, 31.46, 29.65, 28.83, 26.60, 23.03, 22.63, 14.08. HR-MS (FD): calculated for C₅₀H₆₄O₂N₂S₂: 788.44092; found (M⁺): m/z 788.44010.

**(3Z)-6-(benzofuran-2-yl)-3-((Z)-2-(6-(benzofuran-2-yl)-1-(2-butyloctyl)-2-oxoindolin-3-ylidene)ethylidene)-1-(2-butyloctyl)indolin-2-one (EBI-BF-C12):**

**2a** (0.41 g, 0.53 mmol, 1 eq) and trimethyl(benzofuran-2-yl)stannane (0.35 g, 1.26 mmol, 2.4 eq) were placed in a 50 ml dried flask and then, the flask was evacuated and back-filled with argon three times. Next, 15 ml of anhydrous toluene was poured into the flask. A mixture of $Pd_2(dba)_3$ (0.011 g, 0.011 mmol, 0.02 eq) and P($o$-tolyl)$_3$ (12.7 mg) which were separately introduced in 1 ml of anhydrous toluene were injected into the flask. The mixture was heated at 110°C under argon and stirred for 12 hours. Toluene was evaporated. The compound was purified on silica gel chromatography column with hexane and DCM (60:40) as eluent. The product was crystallized in isopropanol three times. 0.27 g of EBI-BF-C$_{12}$ was obtained giving a yield of 55.0%.
[1]H NMR (CDCl$_3$, 300MHz) δ 8.98 (s, 2H), 7.68 (d, J = 7.9 Hz, 2H), 7.59 (d, J = 7.4 Hz, 2H), 7.52 (d, J = 7.6 Hz, 4H), 7.38 – 7.18 (m, 6H), 7.07 (s, 2H), 3.70 (d, J = 5.2 Hz, 4H), 1.97 (s, 2H), 1.46 – 1.17 (m, 32H), 1.01 – 0.74 (m, 12H). [13]C NMR (CDCl$_3$, 300MHz) δ 167.55, 155.50, 154.94, 143.91, 132.05, 130.32, 129.06, 128.86, 124.80, 123.52, 123.14, 121.56, 121.06, 118.71, 111.19, 104.61, 102.55, 44.22, 36.42, 31.87, 31.74, 31.44, 29.68, 28.82, 26.58, 23.06, 22.64, 14.15, 14.10. HR-MS (FD): calculated for $C_{58}H_{68}O_4N_2$: 856.51857; found (M$^+$): m/z 856.51791.

**(3Z)-6-(benzofuran-2-yl)-3-((Z)-2-(6-(benzofuran-2-yl)-1-(2-hexyldecyl)-2-oxoindolin-3-ylidene)ethylidene)-1-(2-hexyldecyl)indolin-2-one (EBI-BF-C16):**

**2b** (0.55 g, 0.61 mmol, 1 eq), trimethyl(benzofuran-2-yl)stannane (0.41 g, 1.46 mmol, 2.4 eq) and P(o-tolyl)$_3$ (14.8 mg, 0.05 mmol, 0.08 eq) were placed in a 50 ml dried flask and then, the flask was evacuated and back-filled with argon three times. Next, 17 ml of anhydrous toluene was poured into the flask. Pd$_2$(dba)$_3$ (0.013 g, 0.012 mmol, 0.02 eq) were separately introduced in 1 ml of anhydrous toluene and were injected into the flask. The mixture was heated at 110°C under argon and stirred for 24 hours. Toluene was evaporated. The compound was purified on silica gel chromatography column with hexane and DCM (50:50) as eluent. The product was crystallized in isopropanol three times. 0.44 g of EBI-BF-C$_{16}$ was obtained giving a yield of 68.1%.
[1]H NMR (TCE-d2, 300MHz): δ 8.94 (s, 2H), 7.67 (d, J = 7.9 Hz, 2H), 7.57 (d, J = 7.5 Hz, 2H), 7.51 (d, J = 5.8 Hz, 4H), 7.36 – 7.18 (m, 6H), 7.08 (s, 2H), 3.65 (d, J = 7.3 Hz, 4H), 1.92 (s, 2H), 1.46 – 1.08 (m, 49H), 0.90 – 0.73 (m, 13H). [13]C NMR (CDCl$_3$, 300MHz) δ 167.51, 155.51, 154.93, 143.88, 132.02, 130.27, 129.06, 128.82, 124.78, 123.52, 123.13, 121.53, 121.04, 118.69, 111.19, 104.59, 102.52, 44.21, 36.41, 31.88, 31.75, 30.00, 29.68, 29.61, 29.32, 26.59, 22.65, 14.09. HR-MS (FD): calculated for $C_{66}H_{84}O_4N_2$: 968.64311; found (M$^+$): m/z 968.64439.

**(3Z)-1-(2-butyloctyl)-3_((Z)-2-(1-(2-butyloctyl)-2-oxo-6-(5-(thiophen-2-yl)thiophen-2-yl)indolin-3-ylidene)ethylidene)-6-(5-(thiophen-2-yl)thiophen-2-yl)indolin-2-one (EBI-2T):**

To a three-necked round bottom flask were added **2a** (0.5 g, 0.64 mmol, 1 eq), 4,4,5,5-tetramethyl-2(5-(thiophen-2-yl)thiophen-2-yl)-1,3,2-dioxaborolane (0.41 g, 1.4 mmol, 2.2 eq), 10 ml of toluene and 10 ml of an aqueous solution of carbonate potassium (2M). The system was placed under argon. Pd(PPh$_3$)$_4$ (74 mg, 0.064 mmol, 0.1 eq) was added. The reaction was heated at 85°C for 24 hours. The product was extracted with DCM and washed with deionized water three times. The organic phase was dried with dehydrated sodium sulfate and DCM was removed under vacuum. The compound was further purified by silica gel chromatography column with chloroform/hexane (60:40) as eluent. 410 mg of EBI-2T were obtained giving a yield of 67.0%.
[1]H NMR (CDCl$_3$, 300MHz): δ 8.82 (s, 2H), 7.51 (d, J= 7.9 Hz, 2H), 7.24-7.15 (m, 6H), 7.13 (d, J=3.3 Hz, 2H), 7.07 (d, J=2.6 Hz, 2H), 6.96 (t, J= 4.0 Hz, 2H), 6.84 (s, 2H), 3.56 (d, J= 6.6 Hz, 4H), 1.82 (s, 2H), 1.43-1.10 (m, 33 H), 0.91-0.68 (m, 12H). [13]C NMR (CDCl$_3$, 300MHz) δ 167.7, 144.03, 142.87, 137.55, 137.19, 135.94, 130.19, 128.60, 127.95, 124.73, 124.70, 124.42, 123.88, 122.75, 121.74, 119.19, 105.36, 44.22, 36.50, 31.88, 31.84, 31.54, 29.69, 28.90, 26.67, 23.07, 22.66, 14.13, 14.09. HR-MS (FD): calculated for $C_{58}H_{68}O_2N_2S_4$: 956.41636; found (M+): m/z 952.41883.

**Devices**

The photovoltaic properties were tested in conventional solar cells following this structures: ITO/PEDOT:PSS/Tri:BTDPP:PC$_{61}$BM/(Ca)/Al. ITO–coated glass substrates were successively cleaned in ultrasonic baths of soap/ DI water x 2 /isopropanol for 10 min, followed by UV-ozone treatment for 15 min. A water solution of PEDOT:PSS, previously filtrated on 0.2 μm, was deposited by spin-coating at 4000 rpm for 60 sec. The layer was dried in the oven at 100 °C under vacuum for 30 min. Then, a solution of EBI with PC$_{61}$BM in chloroform was spin-coated on top of PEDOT:PSS under nitrogen atmosphere. The thickness of the active layer was 100 nm. Calcium (10 nm) and then aluminium (80 nm) were thermally evaporated onto the active layer through shadow masks under 2-4 × 10$^{-6}$ mbar. The effective area was 10.5 mm².

The devices were characterized using a K.H.S SolarCelltest-575 solar simulator with AM 1.5G filters set at 100 mW cm$^{-2}$ with a calibrated radiometer (IL 1400BL). The current density-voltage (J-V) curves measurements were processed with Labview controlled Keithley 2400 SMU. Devices were characterized under nitrogen in a set of glove boxes (O$_2$ and H$_2$O < 0.1 ppm).

Bottom-gate bottom-contact field effect transistors were fabricated (OFETs) using *Fraunhofer IPMS* templates of doped Si with 200 nm-thick silicon oxide and gold electrodes. Substrates were cleaned with successive acetone and IPA baths followed by 15 min of UV-ozone treatment. A solution of 5 mg mL$^{-1}$ of EBI in chloroform was spin-coated on top of transistor substrates in the nitrogen-filled glovebox to form a 40 nm-thick layer. The dimensions of the channel were L= 5 and 10 µm and W= 1 cm. Transistors were measured using an analyzer for semiconductor (Keithley 4200) coupled with a three tip station.

## Acknowledgements

This work was financially supported by the University of Bordeaux and SOLVAY in framework of IDS-FUNMAT network (2012-14-LF). We would like to acknowledge Bertrand Pavageau, Marie-Béatrice Madec and Sabine Goma for their contributions to this work. Jaime Martín acknowledges support from the European Union's Horizon 2020 research and innovation programme under the Marie Skłodowska-Curie grant agreement No. 654682.

## Notes and references